\documentclass[letterpaper,conference]{IEEEtran}

\IEEEoverridecommandlockouts                              % This command is only needed if
\usepackage{amsmath, amssymb,commath,mathtools}

\usepackage{amsthm}
\usepackage{mathbbol}

\usepackage{stfloats}

\usepackage{tikz}
\usetikzlibrary{calc}
\usetikzlibrary{shapes,arrows,fit,patterns,calc,matrix}
\usepackage{graphicx}
\usepackage{dsfont}
\usepackage{changepage}
\usepackage{float,relsize}
\usepackage{textcomp}

\usepackage[margin=.75in]{geometry}

\usepackage{hyperref}
\hypersetup{
    colorlinks,%
    citecolor=black,%
    filecolor=black,%
    linkcolor=black,%
    urlcolor=black,%
    pdfauthor={Molavipour, Bassi, Skoglund},%
    pdftitle={Testing for Directed Information Graphs}%
}

\newcommand{\bb}[1]{\textbf{#1}}
\newcommand{\co}[2]{#1\;\lvert\rvert\; #2}
\DeclareMathOperator*{\argmax}{arg\,max}
\newcommand*{\chaindash}{\,\rule[0.5ex]{1em}{0.55pt}\,}

\newtheorem{definition}{\bf Definition}
\newtheorem{assumption}{\bf Assumption}
\newtheorem{theorem}{\bf Theorem}
\newtheorem{lemma}{\bf Lemma}
\newtheorem{remark}{\bf Remark}
\newtheorem{corollary}{\bf Corollary}

\title{\LARGE \bf
\vspace{.25in}
Testing for Directed Information Graphs
}

\author{Sina Molavipour, Germ\'{a}n Bassi, and Mikael Skoglund
\thanks{This work was supported in part by the Knut and Alice Wallenberg Foundation.}
\thanks{The authors are with the ACCESS Linnaeus Centre, School of Electrical Engineering, KTH Royal Institute of Technology, Stockholm, Sweden (e-mails: {\tt sinmo@kth.se}, {\tt germanb@kth.se}, {\tt skoglund@kth.se}).}%
}

\begin{document}

\maketitle
% \thispagestyle{empty}
% \pagestyle{empty}

%%%%%%%%%%%%%%%%%%%%%%%%%%%%%%%%%%%%%%%%%%%%%%%%%%%%%%%%%%%%%%%%%%%%%%%%%%%%%%%%
\begin{abstract}
In this paper, we study a hypothesis test to determine the underlying directed graph structure of nodes in a network, where the nodes represent random processes and the direction of the links indicate a causal relationship between said processes.
Specifically, a $k$-th order Markov structure is considered for them, and the chosen metric to determine a connection between nodes is the directed information.
The hypothesis test is based on the empirically calculated transition probabilities which are used to estimate the directed information.
For a single edge, it is proven that the detection probability can be chosen arbitrarily close to one, while the false alarm probability remains negligible.
When the test is performed on the whole graph, we derive bounds for the false alarm and detection probabilities, which show that the test is asymptotically optimal by properly setting the threshold test and using a large number of samples.
Furthermore, we study how the convergence of the measures relies on the existence of links in the true graph.    	
\end{abstract}

%%%%%%%%%%%%%%%%%%%%%%%%%%%%%%%%%%%%%%%%%%%%%%%%%%%%%%%%%%%%%%%%%%%%%%%%%%%%%%%%

\section{Introduction}
Causality is a concept that expresses the joint behavior in time of a group of components in a system. In general, it denotes the effect of one component to itself and others in the system during a time period. Consider a network of nodes, each producing a signal in time. These processes can behave independently, or there might be an underlying connection, by nature, between them. Inferring this structure is of great interest in many applications. In~\cite{quinn2011estimating}, for instance, neurons are taken as components while the time series of produced spikes is used to derive the underlying structure. Dynamical models are also a well-known tool to understand functionals of expressed neurons~\cite{friston2003dynamicl}. Additionally, in social networks, there is an increasing interest to estimate influences among users~\cite{quinn2015directed}, while further applications exist in biology~\cite{lord2016inference}, economics~\cite{jiao2013universal}, and many other fields.

Granger~\cite{granger1969investigating} defined the notion of causality between two time series by using a linear autoregressive model and comparing the estimation errors for two scenarios: when history of the second node is accounted for and when it is not. With this definition, however, we can poorly estimate models which operate non-linearly. Directed information was first introduced to address the flow of information in a communication set-up, and suggested by Massey~\cite{massey1990causality} as a measure of causality since it is not limited to linear models. There exist other methods which may qualify for different applications. Several of these definitions are compared in~\cite{quinn2011estimating}, where directed information is argued as a robust measure for causality. There are also symmetric measures like correlation or mutual information, but they can only represent a mutual relationship between nodes and not a directed one.

The underlying causal structure of a network of processes can be properly visualized by a directed graph. In particular, in a Directed Information Graph (DIG) --introduced simultaneously by Amblard and Michel~\cite{Amblard2011} and Quinn \emph{et al.}~\cite{quinn2011estimating}-- the existence of an edge is determined by the value of the directed information between two nodes considering the history of the rest of the network. There are different approaches to tackle the problem of detecting and estimating such type of graphs. Directed information can be estimated based on prior assumptions on the processes' structure, such as Markov properties, and empirically calculating probabilities~\cite{quinn2011estimating, quinn2015directed}. On the other hand, Jiao \emph{et al.}~\cite{jiao2013universal} propose a universal estimator of directed information which is not restricted to any Markov assumption. Nonetheless, in the core of their technique, they consider a context tree weighting algorithm with different depths, which intuitively resembles a learning algorithm for estimating the order of a Markov structure. Other assumptions used in the study of DIGs, that constrain the structure of the underlying graph, are tree structures~\cite{quinn2013efficient} or a limit on the nodes' degree~\cite{quinn2015directed}.
 
The estimation performance on the detection of edges on a DIG is crucial since it allows to characterize, for instance, the optimum test for detection, or the minimum number of samples needed to reliably obtain the underlying model, i.e., the sample complexity. In \cite{quinn2015directed}, the authors derive a bound on the sample complexity using total variation when the directed information between two nodes is empirically estimated. Following that work, Kontoyiannis \emph{et al.}~\cite{kontoyiannis2016estimating} investigate the performance of a test for causality between two nodes, and they show that the convergence rate of the empirical directed information can be improved if it is calculated conditioned on the true relationship between the nodes. In other words, the underlying structure of the true model has an effect on the detection performance of the whole graph. Motivated by this result, in this paper, we study a hypothesis test over a complete graph (not just a link between two nodes) when the directed information is empirically estimated, and we provide interesting insights into the problem. Moreover, we show that for every existing edge in the true graph, the estimation converges with $\mathcal{O}(1/\sqrt{n})$, while if there is no edge in the true model, convergence is of the order of $\mathcal{O}(1/n)$.

The rest of the paper is organized as follows. In Section~\ref{sec:prem}, notations and definitions are introduced. In particular, the directed information is reviewed and the definition of an edge in a DIG is presented. The main results of our work are then shown in Section~\ref{sec:main_result_sec}. First, the performance of a hypothesis test for a single edge is studied, where we analyze the asymptotic behavior of estimators based on the knowledge about the true edges. Then, we demonstrate how the detection of the whole graph relies on the test for each edge. Finally, in the last section, the paper is concluded.

\section{Preliminaries}
\label{sec:prem}
Assume a network with $m$ nodes representing processes $\{\bb{X}_1,\dots,\bb{X}_m\}$. 
The observation of the $l$-th process in the discrete time interval $t_1$ to $t_2$ is described by the random variable $X^{t_2}_{l,t_1}$, which at each time takes values on the discrete alphabet $\mathcal{X}$.
With a little abuse of notation $Y_i$ and $Y_1^n$ represent the observations of the process $\bb{Y}$ at instance $i$ and in the interval $1$ to $n$, respectively.

The metric used to describe the causality relationship of these processes is the directed information, as suggested previously, since it can describe more general structures without further assumptions (such as linearity). Directed information is mainly used in information theory to characterize channels with feedback and it is defined based on \emph{causally} conditioned probabilities.

\begin{definition} The probability distribution of $Y_1^n$ \emph{causally} conditioned on $X_1^n$ is defined as
$$P_{\co{Y_1^n}{X_1^n}}=\prod_{i=1}^n P_{Y_i|X_1^i,Y_1^{i-1}}.$$ 
\end{definition}
\vspace{1mm}

The entropy rate of the process $\bb{Y}$ causally conditioned on $\bb{X}$ is then defined as:
\begin{align}
H(\co{\bb{Y}}{\bb{X}})&\triangleq\lim\limits_{n\to\infty}\frac{1}{n}H(\co{Y_{1}^n}{X_{1}^n})\nonumber\\
&=\lim\limits_{n\to\infty}\frac{1}{n}\sum_{i=1}^{n}H(Y_i|Y^{i-1}_1,X^{i}_1).\nonumber
\end{align}
Consequently, the directed information rate of $\bb{X}$ to $\bb{Y}$ causally conditioned on $\bb{Z}$ is expressed as below:
\begin{align}
I(\co{\bb{X}\to \bb{Y}}{\bb{Z}})&\triangleq H(\co{\bb{Y}}{\bb{Z}})-H(\co{\bb{Y}}{\bb{X},\bb{Z}})\nonumber\\
&=\lim\limits_{n\to\infty}\frac{1}{n}\sum_{i=1}^{n}I(Y_{i}\,;\, X_{1}^{i}|Y_{1}^{i-1},Z_{1}^{i}).
\end{align}

Pairwise directed information does not unequivocally determine the one-step causal influence among nodes in a network. Instead, the history of the other remaining nodes should also be considered. Similarly as introduced in~\cite{quinn2011estimating, Amblard2011}, an edge from $\bb{X}_i$ to $\bb{X}_j$ exists in a directed information graph iff 
\begin{align}\label{edge_existance}
I(\co{\bb{X}_i\to \bb{X}_j}{\bb{X}_{[m]\setminus \{i,j\}}})>0,
\end{align}
where $[m]\triangleq \{1, 2, \ldots, m\}$. Having observed the output of every process, the edges of the graph can be estimated which results in a weighted directed graph.
However, when only the existence of a directed edge is investigated the performance of the detection can be improved. This is presented in Section~\ref{sec:main_result_sec}.

There exist several methods to estimate information theoretic values which most of them intrinsically deal with counting possible events to estimate distributions. One such method is the empirical distribution, which we define as follows.

\begin{definition}
Let $x^n_{[m]}=(x^n_{1,1},x^n_{2,1},\dots,x^n_{m,1})$ be a realization of the random variables $X_{[m]}^n=(X_{1,1}^n,X_{2,1}^n,\dots,X_{m,1}^n)$. The joint empirical distribution of $k'\triangleq k+1$ consecutive time instances of all nodes is then defined as:
\begin{align}\label{emp_dist}
\hat P_{X^{k'}_{[m]}}^{(n)}\!(\mathbb{a}^{k'}_{[m]}) =\frac{1}{n-k}\sum\limits_{t=1}^{n-k} \prod_{i=1}^m \mathds{1}[x^{t+k}_{i,t}=\mathbb{a}^{k'}_{i,1}] ,\quad \forall \mathbb{a}^{k'}_{i,1}\in\mathcal{X}^{k'}.
\end{align}
The joint distribution of any subset of nodes is then a marginal distribution of~\eqref{emp_dist}.
\end{definition}
\vspace{1mm}

By plugging in the empirical distribution we can derive estimators for information theoretic quantities such as the entropy $H$, where we use the notation $\hat H$ to distinguish the empirical estimator, i.e.,
\begin{align}
\hat H(X^{k'})=-\sum_{\mathbb{a}^{k'}\in\mathcal{X}^{k'}}\hat P_{X^{k'}}^{(n)}(\mathbb{a}^{k'}) \log \left(\hat P_{X^{k'}}^{(n)}(\mathbb{a}^{k'}) \right).
\end{align}

A causal influence in the network implies that the past of a group of nodes affects the future of some other group or themselves. This motivates us to focus on a network of joint Markov processes in this paper, since it characterizes a state dependent operation for nodes, although we may put further assumptions to make calculations more tractable. For simplicity, we assume a three-node network (i.e., $m=3$) and the processes to be $\bb{X}, \bb{Y}$, and $\bb{Z}$ in the rest of the work, since the extension of the results for $m>3$ is straightforward.

\vspace{1mm}
\begin{assumption}\label{Markov_assump}
$\{\bb{X},\bb{Y},\bb{Z}\}$ is a jointly stationary Markov process of order $k$. 
\end{assumption}
\vspace{1mm}

Let us define the $\abs{\mathcal{X}}^{3k}\times \abs{\mathcal{X}}^{3}$ transition probability matrix $Q$ with elements  $$Q(X_{k+1},Y_{k+1},Z_{k+1}|X_1^k,Y_1^k,Z_1^k).$$ Then, the next assumption prevents further complexities in the steps of the proof of our main result.

\vspace{1mm}
\begin{assumption}\label{AllPositiv_assump}
All transition probabilities are positive, i.e., $Q>\bf{0}$.
\end{assumption}
\vspace{1mm}

This condition provides ergodicity for the joint Markov process and results in the joint empirical distribution asymptotically converging to the stationary distribution\footnote{The stationary distribution is denoted either as \smash{$P_{\bar X_1^{k+1},\bar Y_1^{k+1},\bar Z_1^{k+1}}$} or as $\bar P_{X_1^{k+1}, Y_1^{k+1}, Z_1^{k+1}}$ in the sequel.%
}, i.e.,
$$\hat P_{X_1^{k+1},Y_1^{k+1},Z_1^{k+1}}^{(n)} \to P_{\bar X_1^{k+1},\bar Y_1^{k+1},\bar Z_1^{k+1}}$$

In general, the directed information rate $I(\co{\bb{X}\to \bb{Y}}{\bb{Z}})$ cannot be expressed with the stationary random variables $\bar X_1^{k+1},\bar Y_1^{k+1}$, and $\bar Z_1^{k+1}$, since a good estimator requires unlimited samples for perfect estimation. To see this,
\begin{align}
I(\co{\bb{X} \to \bb{Y}}{\bb{Z}})&= I(\bar Y_{k+1};\bar X_1^{k+1}|\bar Y_1^k,\bar Z_1^{k+1}) \nonumber\\
&\quad -I(\bar Y_{k+1};\bar Y_{-\infty}^0,\bar Z_{-\infty}^0|\bar Y_1^k,\bar Z_1^{k+1})\nonumber\\
&\leq I(\bar Y_{k+1};\bar X_1^{k+1}|\bar Y_1^k,\bar Z_1^{k+1}),
\label{upperbound_eq}
\end{align}
where we use the Markov property in the first equation, and the inequality holds since the mutual information is non-negative. Thus, with a limited sampling interval, an upper bound would be derived.
The next assumption makes~\eqref{upperbound_eq} hold with equality.

\vspace{1mm}
\begin{assumption}\label{inner_Markov_asuump}
For processes $\bb{Y}$ and $\bb{Z}$, the Markov chain
\begin{align}
\bar Y_{k+1} \chaindash (\bar Y_1^k\bar Z_1^{k+1}) \chaindash (\bar Y_{-\infty}^0\bar Z_{-\infty}^0) \nonumber
\end{align}
holds.
\end{assumption}
\vspace{1mm}

Note that the above assumption should hold for every other two pairs of processes if we are interested in studying the whole graph and not only a single edge.
% In the next section the performance of a hypothesis test is investigated.

\section{Hypothesis test for Directed Information Graphs}
\label{sec:main_result_sec}
Consider a graph $\mathcal{G}$, where the edge from node $i$ to node $j$ is denoted by $v_{ij}$; we say that $v_{ij}=1$ if the node $i$ causally influences the node $j$, otherwise, $v_{ij}=0$. A hypothesis test to identify the graph is performed on the adjacency matrix $V$, whose elements are the $v_{ij}$s, and the performance of the test is studied through its false alarm and detection probabilities
\begin{align}
P_F &=P(\hat V=V^*|V\neq V^*), \label{false alarm}\\
P_D &=P(\hat V=V^*|V=V^*), \label{detection}
\end{align}
where $\hat V$ is the estimation of $V$ (properly defined later), and $V^*$ is the hypothesis model to test. In Theorem~\ref{Main_th} below, both an upper bound on $P_F$ and a lower bound on $P_D$ are derived.

\vspace{1mm}
\begin{theorem}\label{Main_th}
For a directed information graph with adjacency matrix $V$ of size ${m\times m}$, if Assumptions $1$--$3$ hold, the performance of the test for the hypothesis $V^*$ is bounded as:
\begin{align}
P_F &\leq 1-P_G\left(\frac{r}{2},I_{th}\right), \\
P_D &\geq \max\!\left\{1-N_0\!\left[1-P_G\left(\frac{r}{2},I_{th}\right)\right],0\right\}, 
\end{align}
using the plug-in estimation of $n$ samples with $n\to\infty$. The function $P_G$ is the \emph{regularized gamma function}, and $N_0=m(m-1)-N_1$ with $N_1$ denoting the number of directed edges in the hypothesis graph, and $r=\abs{\mathcal{X}}^{mk}( \abs{\mathcal{X}}^{m}-1)$. Finally, $I_{th}$ is the threshold value used to decide the existence of an edge, and its order is $\mathcal{O}(1)$.
\end{theorem}
\vspace{1mm}

The proof of Theorem~\ref{Main_th} consists of two steps. First, the asymptotic behavior of the test for a single edge is derived in Section~\ref{ssec:single_edge_section}. Afterwards, the hypothesis test over the whole graph is studied based on the tests for each single edge. It can be seen later on that, by Remark~\ref{pG_behavior} and Corollary~\ref{corollary_single}, the performance of testing the graph remains as good as a test for causality of a single edge.

\vspace{1mm}
\begin{remark}
\label{pG_behavior}
Note that by increasing $I_{th}$ while remaining of order $\mathcal{O}(1)$, $P_G\left(\frac{r}{2},I_{th}\right)$ gets arbitrarily close to one, which results in the probability of detection converging to one as the probability of false alarm tends to zero. 
\end{remark}
% \vspace{1mm}

\subsection{Asymptotic Behavior of a Single Edge Estimation}\label{ssec:single_edge_section}

In general, every possible probability transition matrix $Q$ can be parametrized with $\theta\in\Theta$, where $\Theta\subset\mathbb{R}^r$ (see Table~\ref{dimension_tab}). The vector $\theta$ is formed by concatenating the elements of $Q$ in a row-wise manner excluding the last (linearly dependent) column. However, if the transition probability could be factorized due to a Markov property among its variables, the matrix might thus be addressed with a lower dimension parameter. 

To see this, let us concentrate in our 3-node network $\{\bb{X},\bb{Y},\bb{Z}\}$. If $v_{xy}=0$, or equivalently $I(\co{\bb{X}\to\bb{Y}}{\bb{Z}})=0$, then by Assumption 3, the transition probability can be factorized as follows,
\begin{flalign}
\label{factorization}
\MoveEqLeft[0]
Q_{\phi_{xy}}(X_{k+1},Y_{k+1},Z_{k+1}|X_1^k,Y_1^k,Z_1^k) = & &\nonumber\\
&Q_{\gamma_{xy}}(X_{k+1},Z_{k+1}|X_1^k,Y_1^k,Z_1^k)Q_{\gamma'_{xy}}(Y_{k+1}|Y_1^k,Z_1^{k+1}). \!\!\!\!\!\!&
\end{flalign}
Here the transition matrix is parametrized by $\phi_{xy}\in\Phi_{xy}$ where $\phi_{xy}$ has two components: $\gamma_{xy}\in\Gamma$ and $\gamma'_{xy}\in\Gamma'$, and $\Phi_{xy}=\Gamma\times\Gamma'$. 
The dimensions of the sets are shown in Table~\ref{dimension_tab}; note that $r>d+d'$.
The vectors $\gamma_{xy}$ and $\gamma'_{xy}$ are also formed by concatenating the elements of their respective matrices as in the case of $\theta$. More details are found in the proof of Theorem~\ref{convergence_theo} in Appendix~\ref{appendix:Proof_hyp_th}.

\begin{table}[!t]
\renewcommand{\arraystretch}{1.3}
\caption{Dimensions of Index Sets for $m=3$.}
\label{dimension_tab}
\centering
\begin{tabular}{|c|c|}
\hline
Set & Dimension\\
\hline
$\Theta$ & $r=\abs{\mathcal{X}}^{3k}( \abs{\mathcal{X}}^{3}-1)$\\
\hline
$\Gamma$ & $d=\abs{\mathcal{X}}^{3k}(\abs{\mathcal{X}}^2 -1)$\\
\hline
$\Gamma'$ & $d'=\abs{\mathcal{X}}^{2k+1}(\abs{\mathcal{X}} -1)$\\
\hline
\end{tabular}
\end{table}

Now consider the Neyman-Pearson criteria to test the hypothesis $\Phi_{xy}$ within $\Theta$.

\vspace{1mm}
\begin{definition}
The log-likelihood is defined as 
\begin{align*}
\MoveEqLeft[1]
L^{\theta}_n(X_1^n,Y_1^n,Z_1^n)\nonumber\\
&=\log\left(Q_{\theta}(X_{k+1}^n,Y_{k+1}^n,Z_{k+1}^n|X_1^k,Y_1^k,Z_1^k) \right)\nonumber\\
&=\log\left( \prod_{i=k+1}^{n} Q_{\theta}(X_i,Y_i,Z_i|X_{i-k}^{i-1},Y_{i-k}^{i-1},Z_{i-k}^{i-1}) \right).
\end{align*}
\end{definition}
\vspace{1mm}

Let $\theta^{\star}$ and $\phi^{\star}_{xy}$ be the most likely choice of transition matrix with general and null hypothesis $v_{xy}=0$, respectively, i.e., 
\begin{align}
\theta^{\star} &= \argmax_{\Theta} L^{\theta}_n(X_1^n,Y_1^n,Z_1^n), \nonumber\\
\phi^{\star}_{xy} &= \argmax_{\Phi_{xy}} L^{\phi_{xy}}_n(X_1^n,Y_1^n,Z_1^n). \label{gamma_defin}
\end{align}
As a result, the test for causality boils down to check the difference between likelihoods, i.e., the log-likelihood ratio:
\begin{align}
\Lambda_{{xy},n}= L^{\theta^{\star}}_n(X_1^n,Y_1^n,Z_1^n)-L^{\phi^{\star}_{xy}}_n(X_1^n,Y_1^n,Z_1^n),
\end{align}
which is the Neyman-Pearson criteria for testing $\Phi_{xy}$ within $\Theta$.
Then, in the following theorem, $\Lambda_{xy,n}$ is shown to converge to a $\chi^2$ distribution of finite degree. The proof follows from standard results in~\cite[Th. 6.1]{billingsley1961statistical}.

\vspace{1mm}
\begin{theorem}\label{convergence_theo}
Consider a network with three nodes $\{\bb{X},\bb{Y},\bb{Z}\}$ and arbitrarily choose two nodes $\bb{X}$ and $\bb{Y}$. Suppose Assumptions~\ref{Markov_assump}--\ref{inner_Markov_asuump} hold, then
$$2\Lambda_{{xy},n}\stackrel{\mathcal{L}}{\to}\chi^2_{r-d-d'},$$
if $v_{xy}=0$ as $n\to \infty$.
\end{theorem}
\begin{IEEEproof}
The conditions of the theorem imply that the true underlying structure for the transition matrix is from $\Phi_{xy}$ which is required as in~\cite[Th. 6.1]{billingsley1961statistical}. The rest of the proof follows similar steps as in~\cite{kontoyiannis2016estimating}. See Appendix~\ref{appendix:Proof_hyp_th} for further details.
\end{IEEEproof}
\vspace{1mm}

\begin{remark}
Note that the asymptotic result from Theorem~\ref{convergence_theo} depends only on the dimensions of the sets and not in the particular pair of nodes involved. Furthermore, the result also holds for a network with more than three nodes by properly defining the dimensions of the sets.
\end{remark}

\vspace{1mm}
\begin{remark}
\label{Remark_th1}
Knowledge about the absence of edges other than $v_{xy}$ in the network results in $\Lambda_{{xy},n}$ converging to a $\chi^2$ distribution of higher degree since~\eqref{factorization} could be further factorized.
To see this, assume $v_{xy}=0$ and consider that a knowledge $S$ about the links (for example, the whole adjacency matrix $V$) was given. Then, let the transition probability be factorized as much as possible, so it can be parametrized by $\Phi'_{xy}$ which has lower or equal dimension than $\Phi_{xy}$. Take
$$\Lambda'_{{xy},n}= L^{\theta^{\star}}_n(X_1^n,Y_1^n,Z_1^n)-L^{\phi'^{\star}_{xy}}_n(X_1^n,Y_1^n,Z_1^n),$$
where $$\phi'^{\star}_{xy}=\argmax_{\Phi'_{xy}} L^{\phi'_{xy}}_n(X_1^n,Y_1^n,Z_1^n).$$
Intuitively, by following similar steps as in the proof of Theorem~\ref{convergence_theo}, we obtain that $\Lambda'_{{xy},n}$ behaves as a $\chi_q^2$ random variable, where $r>q>r-d-d'$. Since the cumulative distribution function of the $\chi_q^2$ is a decreasing function with respect to the degree $q$ then, 
\begin{align}
P_G\!\left(\frac{r}{2},a\right) &\leq P_G\!\left(\frac{q}{2},a\right) \nonumber\\
 &= P(\Lambda'_{xy,n}<a|S, v_{xy}=0) \nonumber\\
 &\leq P(\Lambda_{xy,n}<a|v_{xy}=0) \nonumber\\
 &=P_G\!\left(\frac{r-d-d'}{2},a\right),
 \label{eq:lower_bnd_P}
\end{align}
for sufficiently large $n$ and any $a>0$. The lower bound in~\eqref{eq:lower_bnd_P} allows us to ignore the knowledge about other nodes.
\end{remark}
\vspace{1mm}

Consider now the estimation of the directed information defined as plugging in the empirical distribution (instead of the true distribution) into $I(Y_{k+1}; X_1^{k+1}| Y_1^k,Z_1^{k+1})$, i.e.,
$$\hat I_n^{(k)}(\co{\bb{X}\to\bb{Y}}{\bb{Z}}) \triangleq I(\hat Y_{k+1}; \hat X_1^{k+1}| \hat Y_1^k,\hat Z_1^{k+1}).$$ 
Then, the following lemma states that $\hat I^{(k)}_n(\co{\bb{X}\to\bb{Y}}{\bb{Z}})$, is proportional to $\Lambda_{xy,n}$ with an $\mathcal{O}(n)$ factor. 

\vspace{1mm}
\begin{lemma}\label{lemma_lambda_DI}
$\Lambda_{xy,n}=(n-k)\hat I^{(k)}_n (\co{\bb{X}\to\bb{Y}}{\bb{Z}})$, which is the plug-in estimator of the directed information.
\end{lemma}
\begin{IEEEproof}
The proof follows from standard definitions and noting that the KL-divergence is positive and minimized by zero. See Appendix~\ref{appendix:Proof_lemma_lambda_DI} for the complete proof.
\end{IEEEproof}
\vspace{1mm}

Now, let us define the decision rule for checking the existence of an edge in the graph as follows:
\begin{equation*}
\hat v_{i,j}\triangleq\begin{cases}
1 & \textnormal{if } (n-k)\hat I^{(k)}_n (\co{\bb{X}_i\to\bb{X}_j}{\bb{X}_{[m]\setminus \{i,j\}}})\geq I_{th}\\
0 & \textnormal{o.w.,}
\end{cases}
% \label{threshold_rule}
\end{equation*}
where $I_{th}$ is of order $\mathcal{O}(1)$. Then for any knowledge $S$ about states of edges in the true graph, as long as $v_{xy}=0$ we have:
\begin{align}
\MoveEqLeft[1]
P(\hat v_{xy}=1|S, v_{xy}=0)\nonumber\\
&=P((n-k)\hat I^{(k)}_n (\co{\bb{X}\to\bb{Y}}{\bb{Z}})>I_{th}|S, v_{xy}=0)\nonumber\\
&\leq 1-P_G\!\left(\frac{r}{2},I_{th}\right),
\label{bound_p1}
\end{align}
where the inequality is due to Remark~\ref{Remark_th1}.

From Theorem~\ref{convergence_theo} and Lemma~\ref{lemma_lambda_DI}, it is inferred that when in the true adjacency matrix $v_{xy}=0$, then the empirical estimation of the directed information converges to zero with a $\chi^2$ distribution at a rate $\mathcal{O}(1/n)$.
The asymptotic behavior of $\hat I^{(k)}_n (\co{\bb{X}\to\bb{Y}}{\bb{Z}})$ is different if the edge is present, i.e., $v_{xy}=1$, which is addressed in the following theorem. 

\vspace{1mm}
\begin{theorem}\label{Convergence2_theo}
Consider a network with three nodes $\{\bb{X},\bb{Y},\bb{Z}\}$ and arbitrarily choose two nodes $\bb{X}$ and $\bb{Y}$. Suppose Assumptions~\ref{Markov_assump}--\ref{inner_Markov_asuump} hold and let $\bar I_{xy}\triangleq I(\bar Y_{k+1};\bar X_1^{k+1}|\bar Y_1^k,\bar Z_1^{k+1})$, then,
\begin{equation}
\sqrt{n-k}\left[\hat I^{(k)}_n (\co{\bb{X}\to\bb{Y}}{\bb{Z}})-\bar I_{xy}\right] \to\mathcal{N}(0,\sigma^2),
\end{equation}
with a finite $\sigma^2$ as $n\to \infty$, if $v_{xy}=1$. 
\end{theorem}
\begin{IEEEproof}
The empirical distribution can be decomposed in two parts, where the first one vanishes at a rate faster than $\mathcal{O}(1/\sqrt{n})$ and the second part converges at a rate $\mathcal{O}(1/\sqrt{n})$. The condition $v_{xy}=1$ keeps the second part positive so it determines the asymptotic convergence of $\hat I^{(k)}_n (\co{\bb{X}\to\bb{Y}}{\bb{Z}})$. Refer to Appendix~\ref{appendix:Proof_convergence2_theo} for further details.
\end{IEEEproof}
\vspace{1mm}

\begin{remark}
\label{Remark_th2}
Knowledge about the state of other edges in the true graph model does not affect the asymptotic behavior presented in Theorem~\ref{Convergence2_theo}, given that the condition $v_{xy}=1$ makes the convergence of the estimator independent of all other nodes. This can be seen by following the steps of the proof, where we only use the fact that if the true edge exists then $\bar I_{xy}>0$ and \eqref{factorization} does not hold.
\end{remark}
\vspace{1mm}

We can use Remark~\ref{Remark_th2} to conclude that:
\begin{align}
\MoveEqLeft[1]
P(\hat v_{xy}=0|S, v_{xy}=1)\nonumber\\
&=P((n-k)\hat I^{(k)}_n (\co{\bb{X}\to\bb{Y}}{\bb{Z}})<I_{th}|S, v_{xy}=1) \nonumber \displaybreak[2]\\
&=P((n-k)\hat I^{(k)}_n (\co{\bb{X}\to\bb{Y}}{\bb{Z}})<I_{th}|v_{xy}=1) \nonumber\\
&=1-Q\!\left( \frac{I_{th}-(n-k)\bar I_{xy}}{\sqrt{n-k}\, \sigma} \right), \label{bound_p2}
\end{align}
for sufficiently large $n$, where $Q(\cdot)$ is the Q-function, and where the last equality is due to Theorem~\ref{Convergence2_theo}. Note that if $v_{xy}=1$ then $\bar I_{xy}>0$.

\subsection{Hypothesis Test over an Entire Graph}
\label{ssec:section_graph_hyp}

The performance of testing a hypothesis $V^*$ for a graph is studied by means of the false alarm and detection probabilities defined in~\eqref{false alarm} and~\eqref{detection}, respectively. The results may be considered as an extension of the hypothesis test over a single edge in the graph. 

First, let the false alarm probability be upper-bounded as
\begin{align}
P_F=P(\hat V= V^*|V\neq V^*)\leq\min_{i,j}P(\hat v_{ij}=v^*_{ij}|V\neq V^*).\nonumber
\end{align}
If $V\neq V^*$, there exist nodes $\tau$ and $\rho$ such that $v_{\tau \rho}\neq v^*_{\tau \rho}$. Hence,
\begin{align}
% \MoveEqLeft[1]
P_F &\leq\min_{i,j}P(\hat v_{ij}=v^*_{ij}|V\neq V^*) \label{PF_min_bound1}\\
&\leq P(\hat v_{\tau \rho}=v^*_{\tau \rho}|V\neq V^*)\label{PF_min_bound2}\\
&=P(\hat v_{\tau \rho}=v^*_{\tau \rho}|V\neq V^*, v_{\tau \rho}\neq v^*_{\tau \rho})\nonumber\\
&=\begin{cases}
P(\hat v_{\tau \rho}=0|V\neq V^*, v_{\tau \rho}=1) \\%& \textnormal{if } v_{\tau \rho}=1\\
P(\hat v_{\tau \rho}=1|V\neq V^*, v_{\tau \rho}=0) %& \textnormal{if } v_{\tau \rho}=0
\end{cases}\nonumber\\
&\leq \begin{cases}
1-Q\!\left( \frac{I_{th}-(n-k)\bar I_{\tau \rho}}{\sqrt{n-k}\,\sigma} \right) & \textnormal{if } v_{\tau \rho}=1\\
1-P_G\!\left(\frac{r}{2},I_{th}\right) & \textnormal{if } v_{\tau \rho}=0
\end{cases}\label{PF_bound}
\end{align}
where the last inequality is due to \eqref{bound_p1} and \eqref{bound_p2}.

On the other hand, the complement of the detection probability may be upper-bounded using the union bound:
\begin{align}
&1-P_D=P(\hat V\neq V^*|V= V^*)\leq\sum_{i,j}P(\hat v_{ij}\neq v^*_{ij}|V= V^*)\nonumber \displaybreak[2]\\
&=\sum_{\substack{i,j\\v_{ij}=1}}P(\hat v_{ij}\neq v^*_{ij}|V= V^*)+\sum_{\substack{i,j\\v_{ij}=0}}P(\hat v_{ij}\neq v^*_{ij}|V= V^*) \nonumber\displaybreak[2]\\
&=\sum_{\substack{i,j\\v_{ij}=1}}P(\hat v_{ij}=0|V= V^*,v_{ij}=1)\nonumber\\
&\quad +\sum_{\substack{i,j\\v_{ij}=0}}P(\hat v_{ij}=1|V= V^*,v_{ij}=0)\nonumber\\
%&\leq\sum_{\substack{i,j\\v_{ij}=1}}P(\hat v_{ij}=0|v_{ij}=1)+\sum_{\substack{i,j\\v_{ij}=0}}P(\hat v_{ij}=1|v_{ij}=0), \!\!\!\!
&\leq N_1 \bigg(1-Q\bigg( \frac{I_{th}-(n-k)\bar I}{\sqrt{n-k}\,\sigma} \bigg)\!\! \bigg)+ N_0 \bigg(1-P_G\bigg(\frac{r}{2},I_{th}\bigg)\!\! \bigg),
\label{PD_bound}
\end{align}
where $N_0$ and $N_1$ are the number of off-diagonal $0$s and $1$s in the true matrix $V$, i.e., $N_0+N_1=m(m-1)$, and $\bar I\triangleq\min\limits_{\stackrel{i,j}{\text{s.t.}\, v_{ij}=1}} \bar I_{ij}$. The last inequality holds due to~\eqref{bound_p1} and~\eqref{bound_p2}.

Since $\bar I_{ij}= I(\bar X_{j,k+1};\bar X_{i,1}^{k+1}|\bar X_{j,1}^k,\bar X_{[m]\setminus\{i,j\},1}^{k+1})> 0$ and it is of order $\mathcal{O}(1)$, then as $n\to\infty$ and noting that $$\lim_{a\to\infty}1-Q(-a)=0,$$ we have that,
\begin{align}
 P_F &\leq 1-P_G\!\left(\frac{r}{2},I_{th}\right)\\
 1-P_D &\leq N_0\left[1-P_G\!\left(\frac{r}{2},I_{th}\right)\right].
\end{align}
This concludes the proof of Theorem~\ref{Main_th}.

\vspace{1mm}
\begin{corollary}\label{corollary_single}
In the special case the hypothesis test is performed on a single edge, for the false alarm probability, \eqref{PF_min_bound1} and~\eqref{PF_min_bound2} become equal and we have
\begin{equation*}
P'_F \triangleq P(\hat v_{xy}=1| v_{xy}=0)=1-P_G\left(\frac{r-d-d'}{2},I_{th}\right),
\end{equation*}
and for the detection probability,
\begin{equation*}
 P'_D \triangleq P(\hat v_{xy}=1| v_{xy}=1)=1,
\end{equation*}
as $n\to\infty$, which is in the same lines as the argument in~\cite[Sec. III-C]{kontoyiannis2016estimating} for $m=2$.
\end{corollary}

\begin{figure}[!t]
\centering
\includegraphics[width=\linewidth]{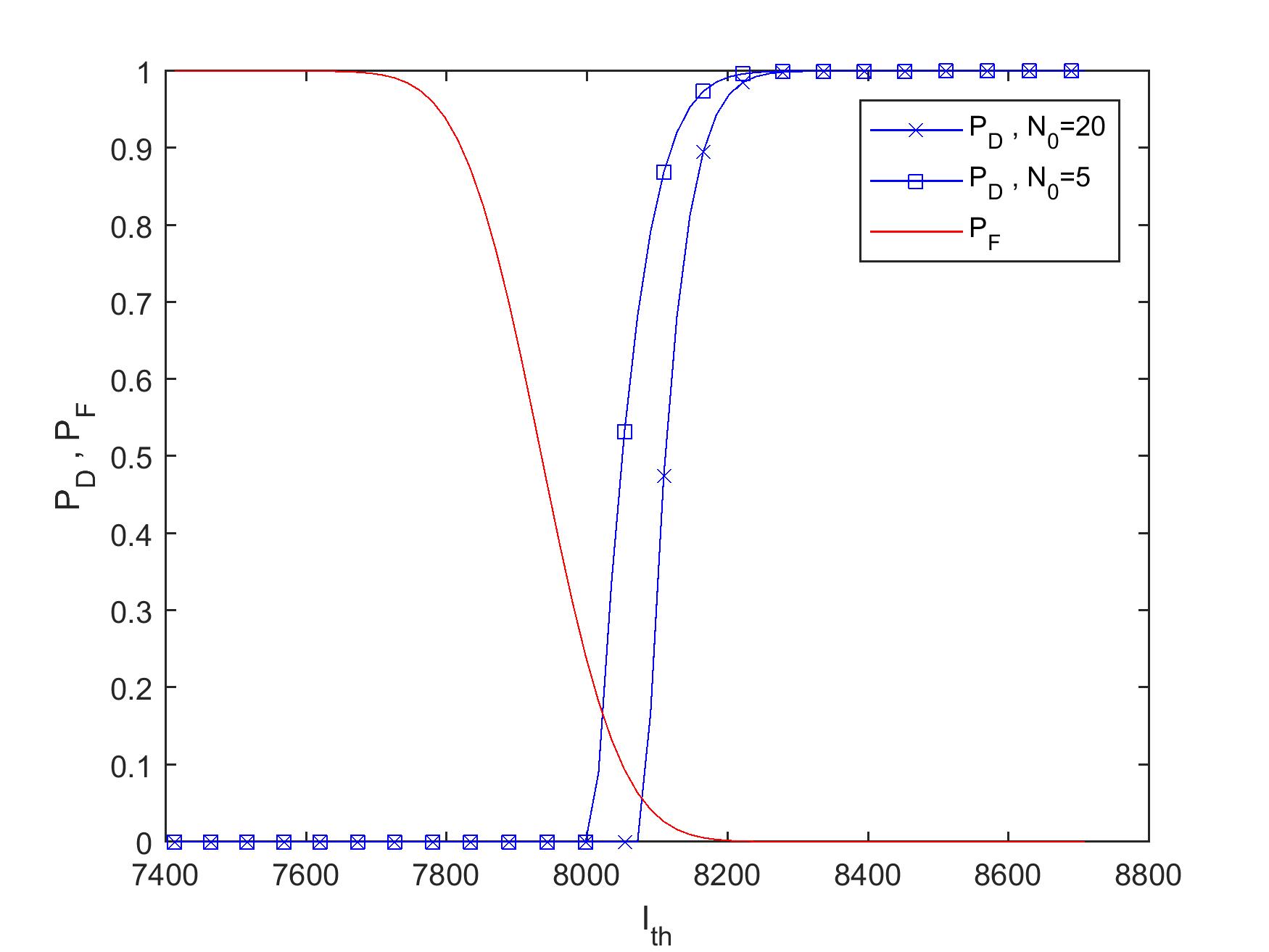}
\caption{Lower bound for detection probability $P_D$ and upper bound for $P_F$, derived by varying the threshold of the test $I_{th}$, and with $k=2$, $m=5$ and binary alphabet. Since $P_F\geq 0$ and $P_D\leq 1$ by increasing $I_{th}$, $P_F\to 0$ and $P_D\to 1$.}
\label{fig_PD}
\end{figure}

\subsection{Numerical Results}

The bounds derived in Theorem~\ref{Main_th} state that the detection probability can be desirably close to one while the false alarm probability can remain near zero with a proper threshold test. In Fig.~\ref{fig_PD}, these bounds are depicted with respect to different values of $I_{th}$ for a network with $m=5$ nodes. The joint process is assumed to be a Markov process of order $k=2$, and the random variables take values on a binary alphabet ($|\mathcal{X}| =2$).

It can be observed in the figure that, for fixed $I_{th}$, $P_D$ improves as $N_0$ decreases, i.e., when the graph becomes sparser. Furthermore, by a proper choice of $I_{th}$, we can reach to optimal performance of the hypothesis test, i.e., $P_D=1$ and $P_F=0$.

\section{Summary and Remarks}
In this paper, we investigated the performance of a hypothesis test for detecting the underlying directed graph of a network of stochastic processes, which represents the causal relationship among nodes, by empirically calculating the directed information as the measure. We showed that the convergence rate of the directed information estimator relies on the existence or not of the link in the real structure. We further showed that with a proper adjustment of the threshold test for single edges, the overall hypothesis test is asymptotically optimal.

This work may be expanded by considering a detailed analysis on the sample complexity of the hypothesis test. Moreover, we assumed in this work that the estimator has access to samples from the whole network while in practice this might not be the case (see e.g.~\cite{scarlett17a}).


%%%%%%%%%%%%%%%%%%%%%%%%%%%%%%%%%%%%%%%%%%%%%%%%%%%%%%%%%%%%%%%%%%%%%%%%%%%%%%%%

%%%%%%%%%%%%%%%%%%%%%%%%%%%%%%%%%%%%%%%%%%%%%%%%%%%%%%%%%%%%%%%%%%%%%%%%%%%%%%%%

%%%%%%%%%%%%%%%%%%%%%%%%%%%%%%%%%%%%%%%%%%%%%%%%%%%%%%%%%%%%%%%%%%%%%%%%%%%%%%%%
\begin{appendices}
\section{Proof of Theorem~\ref{convergence_theo}}
\label{appendix:Proof_hyp_th}

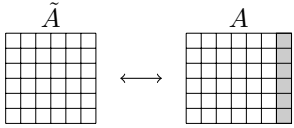
\begin{figure}
\centering
\begin{tikzpicture}
\draw[step=0.2cm,very thin] (-2.4,1.2) grid (-1.2,0);
\filldraw[fill=gray!40, draw=black] (1.2,1.2) rectangle (1.4,0);
\draw[step=0.2cm,very thin] (0,1.2) grid (1.4,0);
\node at (-1.8,1.45) {$\tilde A$};
\node at (0.7,1.4) {$A$};

\node at(-1,0.6) (n1) {};
\node at(-0.2,0.6) (n2) {};
\draw[<->] (n1) to (n2);
\end{tikzpicture}
\caption{The matrix $\tilde{A}$ is formed by removing the last column of $A$.
\label{fig:q_tilde}}
\end{figure}

For any right stochastic matrix $A$ of dimensions $n_a\times m_a$, let the matrix $\tilde A$ denote the first $m_a-1$ linearly independent columns of $A$, as depicted in Fig.~\ref{fig:q_tilde}.

Without loss of generality, consider $\mathcal{X}$ to be the set of integers $\{1, 2, \dots, \abs{\mathcal{X}}\}$ which simplifies the indexing of elements in the alphabet.
Let $u_{x,1}^k$ denote $(u_{x,1},u_{x,2},\dots,u_{x,k})\in\mathcal{X}^k$, and $u'_{x},u'_{y},u'_{z}\in\mathcal{X}$ excluding $(u'_{x},u'_{y},u'_{z})=(\abs{\mathcal{X}},\abs{\mathcal{X}},\abs{\mathcal{X}})$. Next define the $3k+3$ vector
\begin{align}
\vec u\triangleq (u_{x,1}^k,u_{y,1}^k,u_{z,1}^k,u'_x,u'_y,u'_z)
\end{align}
which is associated with an element of $\tilde Q$ (the sub-matrix of the transition probability matrix $Q$).

Every $\vec u$ can be addressed via the pair $(l_{\vec u},g_{\vec u})$ where $l_{\vec u} \in [1:\abs{\mathcal{X}}^{3k}]$ and $g_{\vec u} \in [1:\abs{\mathcal{X}}^{3}-1]$ indicate the row and column of that element, respectively. Also, let $f_{\vec u}\triangleq(l_{\vec u}-1)(\abs{\mathcal{X}}^{3}-1)+g_{\vec u}$, which denotes the index of that element when vectorizing $\tilde{Q}$.
Any possible transition matrix can then be indexed with a vector
$$\theta=(\theta_1,\theta_2,\dots,\theta_r)=(\theta_{f_{\vec u}})\in\Theta$$
as $Q_{\theta}$, where $\Theta$ has dimension $r$ (see Table~\ref{dimension_tab}) and $\theta$ is constructed by concatenation of rows in $\tilde Q_{\theta}$. 

Suppose now that $v_{xy}=0$ or equivalently, by definition \eqref{edge_existance}, $I(\co{\bb{X}\to\bb{Y}}{\bb{Z}})=0$. Then 
\begin{align}\label{factorize2}
&Q(X_{k+1},Y_{k+1},Z_{k+1}|X_1^k,Y_1^k,Z_1^k)\nonumber\\
&= P(X_{k+1},Z_{k+1}|X_1^k,Y_1^k,Z_1^k)P(Y_{k+1}|Y_1^k,Z_1^{k+1}).
\end{align}
Thus, the transition matrix $Q$ is determined by the elements of two other matrices $T_1$ and $T_2$ given by~\eqref{factorize2}. Define the vectors
\begin{align*}
\vec w \,&\triangleq(i_{x,1}^k, i_{y,1}^k, i_{z,1}^k,  i'_x, i'_z), \\
\vec{w}' &\triangleq(i_{y,1}^k, (i_{z,1}^{k}, i'_z), i'_y),
\end{align*}
which are associated with an element in $\tilde T_1$ and $\tilde T_2$, such that $(i'_x,i'_z)\neq(\abs{\mathcal{X}},\abs{\mathcal{X}})$ in $\vec w$ and $i'_y\neq\abs{\mathcal{X}}$ in $\vec {w}'$. Then
\begin{align*}
f_{\vec w} \,&\triangleq(l_{\vec w}-1)(\abs{\mathcal{X}}^{2}-1)+g_{\vec w}, \\
f_{\vec{w}'} &\triangleq(l_{\vec{w}'}-1)(\abs{\mathcal{X}}-1)+g_{\vec{w}'}, 
\end{align*}
where the pairs of row and column indices for each element in $\tilde T_1$ and $\tilde T_2$ are then $(l_{\vec w},g_{\vec w})$ and $(l_{\vec{w}'},g_{\vec{w}'})$, respectively.

A matrix $Q$ such as the one in~\eqref{factorize2} can be parametrized by a vector $\phi_{xy}\in\Phi_{xy}$, where $\Phi_{xy}=\Gamma\times\Gamma'$ has dimension $d\cdot d'$ (see Table~\ref{dimension_tab}). Then
\begin{align*}
& Q_{\phi_{xy}}=\\ 
&\quad  Q_{\gamma_{xy}}(X_{k+1},Z_{k+1}|X_1^k,Y_1^k,Z_1^k)Q_{\gamma'_{xy}}(Y_{k+1}|Y_1^k,Z_1^{k+1}), 
\end{align*}
where
$$\gamma_{xy}=(\gamma_{f_{\vec w}})\in\Gamma\quad \text{and} \quad\gamma'_{xy}=(\gamma'_{f_{\vec{w}'}})\in\Gamma'$$
determine $\phi_{xy}$, are vectors of length $d$ and $d'$, and are constructed by concatenating the rows of $\tilde Q_{\gamma_{xy}}$ and $\tilde Q_{\gamma'_{xy}}$, respectively. 
There exists then the mapping $h:\Phi_{xy}\to\Theta$ such that component-wise: 
\begin{align}\label{mapping_h}
(h(\phi_{xy}))_{f_{\vec u}}=\gamma_{f_{\vec w}} \!\cdot \gamma'_{f_{\vec{w}'}}.
\end{align}

Consider the matrix $K(\phi_{xy})$ of size $(r+1)\times (d+d')$ such that for every element:  
\begin{align}
\label{definition_k}
(K(\phi_{xy}))_{f_{\vec u}, f}=\begin{cases}
\frac{\partial Q_{h(\phi_{xy})}(u'_x,u'_y,u'_z|u_{x,1}^k,u_{y,1}^k,u_{z,1}^k)}{\partial \gamma_{f}}\vspace{0.3cm} & f\leq d\\
\frac{\partial Q_{h(\phi_{xy})}(u'_x,u'_y,u'_z|u_{x,1}^k,u_{y,1}^k,u_{z,1}^k)}{\partial \gamma'_{f-d}} & f>d.
\end{cases}
\end{align}
In other words, every row of the matrix $K(\phi_{xy})$ is a derivative of an element of $Q_{h(\phi_{xy})}$ with respect to all elements of $\gamma_{xy}$ followed by the derivatives with respect to $\gamma'_{xy}$.

According to~\cite[Th. 6.1]{billingsley1961statistical}, and by the definition of $\theta^{\star}$ and $\phi^{\star}_{xy}$ in~\eqref{gamma_defin},
$$2\left\{L^{\theta^{\star}}_n(X_1^n,Y_1^n,Z_1^n)-L^{\phi^{\star}_{xy}}_n(X_1^n,Y_1^n,Z_1^n)\right\}\stackrel{\mathcal{L}}{\to}\chi^2_{r-d-d'},$$
if $Q_{h(\phi_{xy})}$ has continuous third order partial derivatives and $K(\phi_{xy})$ is of rank $d+d'$. 
The first condition holds according to the definition of $h$ in~\eqref{mapping_h}. 
To verify the second condition we can observe that there exist four types of rows in $K(\phi_{xy})$:
\begin{itemize}
\item Type 1: Take the rows $\vec {u}_1=(i_1^k,j_1^k,l_1^k,i',j',l')$ in~\eqref{definition_k} such that $(i',l')\neq(\abs{\mathcal{X}},\abs{\mathcal{X}})$ jointly and $j'\neq\abs{\mathcal{X}}$. This means that in the $(f_{\vec {u}_1})$-th row of $K$, the derivatives are taken from
$$Q_{h(\phi_{xy})}(i',j',l'|i_1^k,j_1^k,l_1^k) = \gamma_{f_{\vec{w}_1}} \!\cdot \gamma'_{f_{\vec{w}'_1}}, $$
where $\vec {w}_1=(i_1^k,j_1^k,l_1^k,i',l')$ and $\vec {w}'_1=(j_1^k, (l_1^{k}, l'), j')$.
So all elements in such rows are zero except in the columns $f_{\vec{w}_1}$ and $(d+f_{\vec{w}'_1})$, which take the values $\gamma'_{f_{\vec{w}'_1}}$ and $\gamma_{f_{\vec{w}_1}}$, respectively.

\item Type 2: Now consider the rows $\vec{u}_2=(i_1^k,j_1^k,l_1^k,i',j',l')$ such that $(i',l')=(\abs{\mathcal{X}},\abs{\mathcal{X}})$ and $j'\neq\abs{\mathcal{X}}$. This means that in the $(f_{\vec {u}_2})$-th row of $K$, the derivatives are taken from 
\begin{multline*}
Q_{h(\phi_{xy})}(i',j',l'|i_1^k,j_1^k,l_1^k) \\
=\bigg(1-\sum_{(a,b)\neq(\abs{\mathcal{X}},\abs{\mathcal{X}})}\gamma_{f_{\vec{ w}_2(a,b)}}\bigg) \gamma'_{f_{\vec{ w}'_2}},
\end{multline*}
where we define $\vec {w}_2(a,b)=(i_1^k,j_1^k,l_1^k,a,b)$ and $\vec {w}'_2=(j_1^k, (l_1^{k}, l'), j')$.
So all elements in such rows are zero except in the $(\abs{\mathcal{X}}^2-1)$ columns (among the first $d$ columns) from $f_{\vec{ w}_2(1,1)}$ to $f_{\vec{ w}_2(\abs{\mathcal{X}}, \abs{\mathcal{X}}-1)} $ which are equal to $-\gamma'_{f_{\vec{ w}'_2}}$, and the column $(d+f_{\vec{w}'_2})$ which is equal to $$1-\sum_{(a,b)\neq(\abs{\mathcal{X}},\abs{\mathcal{X}})}\gamma_{f_{\vec{ w}_2(a,b)}}.$$

\item Type 3: Consider the rows $\vec{u}_3 =(i_1^k,j_1^k,l_1^k,i',j',l')$ such that $(i',l')\neq(\abs{\mathcal{X}},\abs{\mathcal{X}})$ and $j'=\abs{\mathcal{X}}$. Also let $\vec {w}_3=(i_1^k,j_1^k,l_1^k,i',l')$ and $\vec{w}'_3(a)=(j_1^k, (l_1^{k}, l'), a)$. Then, all elements of such rows are zero except in the column $f_{\vec{w}_3}$ which takes the value
$$1-\sum_{a\neq\abs{\mathcal{X}}}\gamma'_{f_{\vec{w}'_3(a)}},$$
and the $(\abs{\mathcal{X}}-1)$ columns (among the last $d'$ columns) from $d+f_{\vec{w}'_3(1)}$ to $d+f_{\vec{w}'_3(\abs{\mathcal{X}}-1)}$ that are equal $-\gamma_{f_{\vec{w}_3}}$.

\item Type 4: Lastly, consider rows $\vec{u}_4=(i_1^k,j_1^k,l_1^k,i',j',l')$ such that $(i',l')=(\abs{\mathcal{X}},\abs{\mathcal{X}})$ and $j'=\abs{\mathcal{X}}$. Assume vectors $\vec {w}_4(a,b)=(i_1^k,j_1^k,l_1^k,a,b)$ and $\vec{w}'_4(a)=(j_1^k, (l_1^{k}, l'), a)$. Then , the only non-zero elements belong to the $(\abs{\mathcal{X}}^2-1)$ columns from $f_{\vec{w}_4(1,1)}$ to $f_{\vec{w}_4(\abs{\mathcal{X}},\abs{\mathcal{X}})}$ (among the first $d$ columns) which are equal to $$-\bigg(1-\sum_{a\neq\abs{\mathcal{X}}}\gamma'_{f_{\vec{w}'_4(a)}}\bigg),$$
and the $(\abs{\mathcal{X}}-1)$ columns from $d+f_{\vec{w}'_4(1)}$ to $d+f_{\vec{w}'_4(\abs{\mathcal{X}}-1)}$ (among the last $d'$ columns) which are equal to
$$-\bigg(1-\sum_{(a,b)\neq(\abs{\mathcal{X}},\abs{\mathcal{X}})}\gamma_{f_{\vec{ w}_4(a,b)}}\bigg).$$
\end{itemize}

We show now that if a linear combination of all columns equals the vector zero, then all coefficients should be zero as well. Let $c_f$ be the $f$-th column of $K(\phi_{xy})$ then if 
\begin{align}
\label{linear_comb_col}
\sum_{f=1}^{d+d'}\alpha_f c_f=\vec 0,
\end{align}
then, $\alpha_f=0, \forall f$. To see this, consider the Type 1 row with $i_1^k=j_1^k=l_1^k=1^k$ and $i'=l'=1$. Since it only has two non-zero elements, we have that
\begin{equation}
\forall j'\in[1:\abs{\mathcal{X}}-1]:\quad \alpha_{1}\gamma'_{j'}+\alpha_{j'}\gamma_{1} = 0.
\label{sign1}
\end{equation}
Then, take the Type 3 row with $i_1^k=j_1^k=l_1^k=1^k$ and $i'=l'=j'=1$, where we have that
\begin{equation}
\alpha_{1}\bigg(\sum_{a\neq\abs{\mathcal{X}}} 1-\gamma'_{a}\bigg)-\sum_{b\neq\abs{\mathcal{X}}}\alpha_{b}\gamma_{1} = 0.
\label{sign2}
\end{equation}
From~\eqref{sign1} and noting that we have assumed $Q>\bf{0}$, if $\alpha_1>0$ then $\alpha_{j'}<0$ for all $j'\in[1:\abs{\mathcal{X}}-1]$. Hence, the left-hand side of~\eqref{sign2} is strictly positive and not zero. An analogous result is found assuming $\alpha_1<0$. By contradiction, we conclude that $\alpha_1=0$, and from~\eqref{sign1}, 
$$\forall j'\in[1:\abs{\mathcal{X}}-1]:\quad \alpha_{j'}=0.$$ 
By varying $(i',l')$ and for all combinations of $(i_1^k,j_1^k,l_1^k)$ we derive that all $\alpha_f$s are zero, and as a result, $K(\phi_{xy})$ has $d+d'$ linearly independent columns which meets the second condition.
The proof of Theorem~\ref{convergence_theo} is thus complete.
\endIEEEproof

\section{Proof of Lemma~\ref{lemma_lambda_DI}}
\label{appendix:Proof_lemma_lambda_DI}

The proof follows similar steps as the one in~\cite[Prop.~9]{kontoyiannis2016estimating}. Using the definition of log-likelihood,
\begin{align}
& L^{\theta^{\star}}_n(X_1^n,Y_1^n,Z_1^n)\nonumber\\
&=\max\limits_{\theta\in\Theta}\sum_{i=k+1}^{n} \log(Q_{\theta}(X_i,Y_i,Z_i|X_{i-k}^{i-1},Y_{i-k}^{i-1},Z_{i-k}^{i-1}))\nonumber\\
& =\max\limits_{\theta\in\Theta}\sum_{x_1^{k+1}y_1^{k+1}z_1^{k+1}} (n-k)\hat P_{X_1^{k+1}Y_1^{k+1}Z_1^{k+1}}(x_1^{k+1}y_1^{k+1}z_1^{k+1})\nonumber\\
&\qquad\times\log(Q_{\theta}(x_{k+1}y_{k+1}z_{k+1}|x_1^{k}y_1^{k}z_1^{k}))\nonumber\\
&=-(n-k)\biggl[\min\limits_{\theta\in\Theta}\biggl\{D\big(\co{\hat P_{X_1^{k+1}Y_1^{k+1}Z_1^{k+1}}}{Q_{\theta}\otimes \hat P_{X_1^kY_1^kZ_1^k}}\big)\biggr\}\nonumber\\
&\quad + \!\!\!\!\! \sum_{x_1^{k+1}y_1^{k+1}z_1^{k+1}} \!\!\!\!\!\! \hat P(x_1^{k+1}y_1^{k+1}z_1^{k+1})\log\bigg(\frac{\hat P(x_1^{k}y_1^{k}z_1^{k})}{\hat P(x_1^{k+1}y_1^{k+1}z_1^{k+1})}\bigg)\biggl],
\label{loglikelihood_expansion}
\end{align}
 where 
\begin{multline*}
(Q_{\theta}\otimes \hat P_{X_1^kY_1^kZ_1^k})(x_1^{k+1}y_1^{k+1}z_1^{k+1})\triangleq\\
\qquad \hat P_{X_1^kY_1^kZ_1^k}(x_1^{k}y_1^{k}z_1^{k})Q_{\theta}(x_{k+1}y_{k+1}z_{k+1}|x_1^{k}y_1^{k}z_1^{k}).
\end{multline*}
Since the KL-divergence is minimized by zero, then
\begin{multline}
L^{\theta^{\star}}_n(X_1^n,Y_1^n,Z_1^n) =(n-k)\cdot \\
 [\hat H(X_1^k,Y_1^k,Z_1^k) - \hat H(X_1^{k+1},Y_1^{k+1},Z_1^{k+1})].
 \label{likelihood1}
\end{multline}

On the other hand, for the second log-likelihood, we have:
\begin{align}
\MoveEqLeft[1]
L^{\phi^{\star}}_n(X_1^n,Y_1^n,Z_1^n)\nonumber\\
&=\max\limits_{\phi\in\Phi}\sum_{i=k+1}^{n} \log(Q_{\phi}(X_i,Y_i,Z_i|X_{i-k}^{i-1},Y_{i-k}^{i-1},Z_{i-k}^{i-1})) \nonumber \displaybreak[2]\\
&=\underbrace{\max\limits_{\phi^{xz}}\sum_{i=k+1}^{n} \log(Q_{\phi^{xz}}(X_i,Z_i|X_{i-k}^{i-1},Y_{i-k}^{i-1},Z_{i-k}^{i-1}))}_{A_1}
\nonumber\\
&\quad+\underbrace{\max\limits_{\phi^{y}}\sum_{i=k+1}^{n}\log(Q_{\phi^y}(Y_i|Y_{i-k}^{i-1},Z_{i-k}^{i}))}_{A_2}.\nonumber
\end{align}
With a similar approach as in~\eqref{loglikelihood_expansion}, we can expand $A_1$ and $A_2$ as it is shown in~\eqref{eq:expan_a1a2} at the bottom of the page.
\begin{figure*}[!b]
\normalsize
\hrulefill
\vspace*{4pt}
\begin{align}
A_1 &= 
 -(n-k)\biggl[\min\limits_{\phi^{xz}}\biggl\{D\big(\co{\hat P_{X_1^{k+1},Y_1^{k},Z_1^{k+1}}}{Q_{\phi^{xz}}\otimes \hat P_{X_1^{k},Y_1^{k},Z_1^{k}}}\big)\biggr\} 
 + \!\!\! \sum_{x_1^{k+1},y_1^{k},z_1^{k+1}} \!\!\! \hat P(x_1^{k+1},y_1^{k},z_1^{k+1})\log\bigg(\frac{\hat P(x_1^{k},y_1^{k},z_1^{k})}{\hat P(x_1^{k+1},y_1^{k},z_1^{k+1})}\bigg)\biggr] \nonumber\\
A_2 &= 
 -(n-k)\biggl[\min\limits_{\phi^{y}}\biggl\{D\big(\co{\hat P_{Y_1^{k+1},Z_1^{k+1}}}{Q_{\phi^{y}}\otimes\hat P_{Y_1^{k},Z_1^{k+1}}}\big)\biggr\} 
 +\sum_{y_1^{k+1},z_1^{k+1}} \hat P(y_1^{k+1},z_1^{k+1})\log\bigg(\frac{\hat P(y_1^{k},z_1^{k+1})}{\hat P(y_1^{k+1},z_1^{k+1})}\bigg)\biggr].
\label{eq:expan_a1a2}
\end{align}
\end{figure*}
As a result,
\begin{align}
\MoveEqLeft[1]
L^{\phi^{\star}}_n(X_1^n,Y_1^n,Z_1^n) \nonumber\\
 &= (n-k)\Bigl[\hat H(X_1^{k},Y_1^{k},Z_1^{k}) -\hat H(X_1^{k+1},Y_1^{k},Z_1^{k+1}) \nonumber\\
 &\quad  + \hat H(Y_1^{k},Z_1^{k+1})-\hat H(Y_1^{k+1},Z_1^{k+1}) \Bigr].
\label{likelihood2}
\end{align}

Finally, combining~\eqref{likelihood1} and~\eqref{likelihood2}, we obtain
\begin{align}
&\Lambda_{{xy},n}= L^{\theta^{\star}}_n(X_1^n,Y_1^n,Z_1^n)-L^{\phi^{\star}_{xy}}_n(X_1^n,Y_1^n,Z_1^n)\nonumber\\
% &=(n-k)[\hat H(X_1^{k+1},Y_1^{k},Z_1^{k+1})-\hat H(X_1^{k+1},Y_1^{k+1},Z_1^{k+1})\nonumber\\
% &\hspace{3.55cm}- \hat H(Y_1^{k},Z_1^{k+1})+\hat H(Y_1^{k+1},Z_1^{k+1})]\nonumber\\
&=(n-k)[\hat H(Y_{k+1}|Y_1^k,Z_1^{k+1})-\hat H(Y_{k+1}|X_1^{k+1},Y_1^{k},Z_1^{k+1})]\nonumber\\
&=(n-k)\hat I_n^{(k)}(\co{\bb{X}\to\bb{Y}}{\bb{Z}}),
\end{align}
which concludes the proof of Lemma~\ref{lemma_lambda_DI}.
\endIEEEproof

% ==================
% This equation is for next section but it should appear before!
\begin{figure*}[!b]
\vspace*{-10pt}
\normalsize
\hrulefill
\vspace*{4pt}
\begin{align}
\MoveEqLeft[1]
\hat I^{(k)}_n (\co{\bb{X}\to\bb{Y}}{\bb{Z}}) = \sum\nolimits_{\mathbb{x}_1^{k+1} \mathbb{y}_1^{k+1} \mathbb{z}_1^{k+1}} \hat P(\mathbb{x}_1^{k+1} \mathbb{y}_1^{k+1} \mathbb{z}_1^{k+1}) \log\frac{\hat P(\mathbb{y}_{k+1} \mathbb{x}_1^{k+1}|\mathbb{y}_1^k \mathbb{z}_1^{k+1})}{\hat P(\mathbb{y}_{k+1}|\mathbb{y}_1^k \mathbb{z}_1^{k+1})\hat P(\mathbb{x}_1^{k+1}|\mathbb{y}_1^k \mathbb{z}_1^{k+1})}\nonumber\\[3pt]
&=\sum\nolimits_{\mathbb{x}_1^{k+1} \mathbb{y}_1^{k+1} \mathbb{z}_1^{k+1}} \ \hat P(\mathbb{x}_1^{k+1} \mathbb{y}_1^{k+1} \mathbb{z}_1^{k+1})\log\biggl[\frac{\hat P(\mathbb{y}_{k+1} \mathbb{x}_1^{k+1}|\mathbb{y}_1^k \mathbb{z}_1^{k+1})\bar P(\mathbb{y}_{k+1}|\mathbb{y}_1^k \mathbb{z}_1^{k+1})\bar P(\mathbb{x}_1^{k+1}|\mathbb{y}_1^k \mathbb{z}_1^{k+1})}{\hat P(\mathbb{y}_{k+1}|\mathbb{y}_1^k \mathbb{z}_1^{k+1})\hat P(\mathbb{x}_1^{k+1}|\mathbb{y}_1^k \mathbb{z}_1^{k+1})\bar P(\mathbb{y}_{k+1} \mathbb{x}_1^{k+1}|\mathbb{y}_1^k \mathbb{z}_1^{k+1})}\biggr]\nonumber\\
&\quad+\sum\nolimits_{\mathbb{x}_1^{k+1} \mathbb{y}_1^{k+1} \mathbb{z}_1^{k+1}}\hat P(\mathbb{x}_1^{k+1} \mathbb{y}_1^{k+1} \mathbb{z}_1^{k+1}) \log\frac{\bar P(\mathbb{y}_{k+1} \mathbb{x}_1^{k+1}|\mathbb{y}_1^k \mathbb{z}_1^{k+1})}{\bar P(\mathbb{y}_{k+1}|\mathbb{y}_1^k \mathbb{z}_1^{k+1})\bar P(\mathbb{x}_1^{k+1}|\mathbb{y}_1^k \mathbb{z}_1^{k+1})}\nonumber\\[3pt]
&=\sum\nolimits_{\mathbb{x}_1^{k+1} \mathbb{y}_1^{k+1} \mathbb{z}_1^{k+1}}\hat P(\mathbb{x}_1^{k+1} \mathbb{y}_1^{k+1} \mathbb{z}_1^{k+1})\log\biggl[\frac{\hat P(\mathbb{x}_1^{k+1} \mathbb{y}_1^{k+1} \mathbb{z}_1^{k+1}) \hat P(\mathbb{y}_1^k \mathbb{z}_1^{k+1})}{\hat P(\mathbb{y}_1^{k+1} \mathbb{z}_1^{k+1})\hat P(\mathbb{x}_1^{k+1} \mathbb{y}_1^k \mathbb{z}_1^{k+1})} \cdot \frac{\bar P(\mathbb{y}_1^{k+1} \mathbb{z}_1^{k+1})\bar P(\mathbb{x}_1^{k+1} \mathbb{y}_1^k \mathbb{z}_1^{k+1})}{\bar P(\mathbb{x}_1^{k+1} \mathbb{y}_1^{k+1} \mathbb{z}_1^{k+1}) \bar P(\mathbb{y}_1^k \mathbb{z}_1^{k+1})}\biggl]\nonumber\\
&\quad+\sum\nolimits_{\mathbb{x}_1^{k+1} \mathbb{y}_1^{k+1} \mathbb{z}_1^{k+1}}\hat P(\mathbb{x}_1^{k+1} \mathbb{y}_1^{k+1} \mathbb{z}_1^{k+1}) \log\frac{\bar P(\mathbb{y}_{k+1} \mathbb{x}_1^{k+1}|\mathbb{y}_1^k \mathbb{z}_1^{k+1})}{\bar P(\mathbb{y}_{k+1}|\mathbb{y}_1^k \mathbb{z}_1^{k+1})\bar P(\mathbb{x}_1^{k+1}|\mathbb{y}_1^k z_1^{k+1})}\nonumber\\[3pt]
&=D(\co{\hat P_{X_1^{k+1} Y_1^{k+1} Z_1^{k+1}}}{\bar P_{X_1^{k+1} Y_1^{k+1} Z_1^{k+1}}}) + 
D(\co{\hat P_{Y_1^k Z_1^{k+1}}}{\bar P_{Y_1^k Z_1^{k+1}}})\nonumber\\[3pt]
&\quad-D(\co{\hat P_{Y_1^{k+1} Z_1^{k+1}}}{\bar P_{Y_1^{k+1} Z_1^{k+1}}})-D(\co{\hat P_{X_1^{k+1} Y_1^k Z_1^{k+1}}}{\bar P_{X_1^{k+1} Y_1^k Z_1^{k+1}}})\nonumber\\ 
&\quad+ \sum_{\mathbb{x}_1^{k+1} \mathbb{y}_1^{k+1} \mathbb{z}_1^{k+1}}\left(\frac{1}{n-k}\sum\nolimits_{i=k+1}^{n}\mathds{1}[x_{i-k}^{i} y_{i-k}^{i} z_{i-k}^{i}=\mathbb{x}_1^{k+1} \mathbb{y}_1^{k+1} \mathbb{z}_1^{k+1}]\right)\log\frac{\bar P(\mathbb{y}_{k+1} \mathbb{x}_1^{k+1}|\mathbb{y}_1^k \mathbb{z}_1^{k+1})}{\bar P(\mathbb{y}_{k+1}|\mathbb{y}_1^k \mathbb{z}_1^{k+1})\bar P(\mathbb{x}_1^{k+1}|\mathbb{y}_1^k \mathbb{z}_1^{k+1})}.\label{DI_expansion}
\end{align}
%\vspace{.5cm}
%\hrulefill
%\begin{align}
%&D(\co{\hat P_{X_1^{k+1} Y_1^{k+1} Z_1^{k+1}}}{\bar P_{X_1^{k+1} Y_1^{k+1} Z_1^{k+1}}})=\log\log (n-k)\sum_{x_1^{k+1} y_1^{k+1} z_1^{k+1}}{\left(\hat P(x_1^{k+1} y_1^{k+1} z_1^{k+1})-\bar P(x_1^{k+1} y_1^{k+1} z_1^{k+1})\right)^2}\frac{C}{\log\log (n-k)}\label{expansion_D_seq}
%\end{align}
% Restore the current equation number.
%\setcounter{equation}{\value{MYtempeqncnt}}
% IEEE uses as a separator
\end{figure*}

% \addtolength{\textheight}{-2.5cm}

\section{Proof of Theorem~\ref{Convergence2_theo}}
\label{appendix:Proof_convergence2_theo}

We begin by expanding the expression $\hat I^{(k)}_n (\co{\bb{X}\to\bb{Y}}{\bb{Z}})$ using the definition of the empirical distribution in~\eqref{emp_dist} and we obtain~\eqref{DI_expansion}, found at the bottom of the next page. We then proceed to analyze the asymptotic behavior of the estimator.
% 
% The equation was placed before this section so it would appear in the correct page.
% 

The first four terms in~\eqref{DI_expansion}, i.e., the KL-divergence terms, decay faster than $\mathcal{O}(1/\sqrt{n})$. This is shown later in the proof. On the other hand, since $v_{xy}=1$, $I(\bar Y_{k+1};\bar X_{1}^{k+1}|\bar Y_{1}^{k},\bar Z_{1}^{k+1})>0$ due to~\eqref{upperbound_eq} and Assumption~\ref{inner_Markov_asuump}, and thus, the last term in~\eqref{DI_expansion} is non-zero and dominates the convergence of the estimator, as we see next. Here, one observes that conditioning on $v_{xy}=1$ is sufficient to analyze the convergence of $\hat I^{(k)}_n (\co{\bb{X}\to\bb{Y}}{\bb{Z}})$ and further knowledge about other edges is irrelevant (see Remark~\ref{Remark_th2}). We then conclude that,
\begin{equation*}
\lim_{n\to\infty} \sqrt{n-k}\, \hat I^{(k)}_n (\co{\bb{X}\to\bb{Y}}{\bb{Z}}) = \lim_{n\to\infty} \frac{1}{\sqrt{n-k}}\sum_{i=k+1}^{n}S_i,
\end{equation*}
where
\begin{align*}
S_i &\triangleq\log P_{\bar Y_{k+1} \bar X_1^{k+1}|\bar Y_1^k \bar Z_1^{k+1}}(y_i x_{i-k}^i|y^{i-1}_{i-k} z_{i-k}^i) \\
&\quad - \log P_{\bar Y_{k+1}|\bar Y_1^k \bar Z_1^{k+1}}(y_i|y^{i-1}_{i-k} z_{i-k}^i) \\
&\quad -\log P_{\bar X_1^{k+1}|\bar Y_1^k \bar Z_1^{k+1}}(x_{i-k}^i|y^{i-1}_{i-k} z_{i-k}^i).
\end{align*}
We note that $S_i$ is a functional of the chain $\{(X_{i-k}^i, Y_{i-k}^i, \allowbreak Z_{i-k}^i)\}$ and its mean is $\mathds{E}[S]=I(\bar Y_{k+1};\bar X_{1}^{k+1}|\bar Y_{1}^{k},\bar Z_{1}^{k+1})$. The chain is ergodic and we can thus apply the central limit theorem~\cite[Sec.~I.16]{chung2012markov} to the partial sums to obtain
\begin{align}
\sqrt{n-k}\left(\frac{1}{n-k}\sum_{i=k+1}^{n-1}S_i - \mathds{E}[S]\right)\to \mathcal{N}(0,\sigma^2),
\end{align}
where $\sigma^2$ is bounded.

Now, to complete the proof, it only remains to show that the KL-divergence terms in~\eqref{DI_expansion} multiplied by a $\sqrt{n-k}$ factor converge to zero as $n\to\infty$. We present the proof for one term and the others follow a similar approach. We first recall the Taylor expansion with Lagrange remainder form, 
\begin{equation*}
f(x)=f(a)+f'(a)(x-a) +\frac{f''(x^*)(x-a)^2}{2!},
\end{equation*}
for some $x^*\in (a,x)$. Then, let us define $\rho\triangleq \frac{\bar P(x_1^{k+1} y_1^{k+1} z_1^{k+1})}{\hat P(x_1^{k+1} y_1^{k+1} z_1^{k+1})},$ so we can expand the first KL-divergence term as:
\begin{flalign}
\MoveEqLeft[0.5]
D(\co{\hat P_{X_1^{k+1} Y_1^{k+1} Z_1^{k+1}}}{\bar P_{X_1^{k+1} Y_1^{k+1} Z_1^{k+1}}}) &\nonumber\\
&= -\sum\nolimits_{x_1^{k+1} y_1^{k+1} z_1^{k+1}}\hat P(x_1^{k+1} y_1^{k+1} z_1^{k+1})\log\rho &\nonumber\\
&= -\sum\hat P(x_1^{k+1} y_1^{k+1} z_1^{k+1}) \Big[(\rho-1) - \frac{(\rho-1)^2}{2!\tau^2} \Big] &\nonumber\\
&= \sum\hat P(x_1^{k+1} y_1^{k+1} z_1^{k+1})(\rho-1)^2\frac{1}{2\tau^2} &\label{prob_sum}\\
&= \sum\left(\hat P(x_1^{k+1} y_1^{k+1} z_1^{k+1})-\bar P(x_1^{k+1} y_1^{k+1} z_1^{k+1})\right)^2C, \!\!\!\!\!&\label{klExpan}
\end{flalign}
for some $\tau\in (1,\rho)$, where
$$C\triangleq\frac{1}{2\hat P(x_1^{k+1} y_1^{k+1} z_1^{k+1})\tau^2},$$
and~\eqref{prob_sum} follows due to 
\begin{align*}
\MoveEqLeft[1]
\sum\hat P(x_1^{k+1} y_1^{k+1} z_1^{k+1})(\rho-1)\\
&=\sum\bar P(x_1^{k+1} y_1^{k+1} z_1^{k+1})-\hat P(x_1^{k+1} y_1^{k+1} z_1^{k+1})=0.
\end{align*} 
Since the Markov model is assumed to be ergodic (Assumption~\ref{AllPositiv_assump}), $\hat P(x_1^{k+1} y_1^{k+1} z_1^{k+1})\nrightarrow 0$, and therefore $C$ is bounded. 
Now $\forall i\in[1:n-k]$ consider the sequence 
$$T_i(x_1^{k+1} y_1^{k+1} z_1^{k+1})\triangleq\mathds{1}[X_i^{k+i} Y_i^{k+i} Z_i^{k+i}=x_1^{k+1} y_1^{k+1} z_1^{k+1}]$$
with mean $\bar P(x_1^{k+1} y_1^{k+1} z_1^{k+1})$.
According to the law of iterated logarithms,
$$ \limsup_{n\to\infty}\frac{\sum_{i=1}^{n-k}(T_i - \bar P(x_1^{k+1} y_1^{k+1} z_1^{k+1}))}{\sqrt{(n-k)\log\log(n-k)}}=\sqrt{2} \quad \textnormal{a.s.}$$
Using the definition of the empirical distribution, this implies
\begin{align}
\MoveEqLeft[1]
\limsup_{n\to\infty}\frac{(n-k)(\hat P(x_1^{k+1} y_1^{k+1} z_1^{k+1})-\bar P(x_1^{k+1} y_1^{k+1} z_1^{k+1}))}{\sqrt{(n-k)\log\log(n-k)}}\nonumber\\
 &=\limsup_{n\to\infty}\frac{\hat P(x_1^{k+1} y_1^{k+1} z_1^{k+1})-\bar P(x_1^{k+1} y_1^{k+1} z_1^{k+1})}{\sqrt{\log\log(n-k)}/\sqrt{n-k}}\nonumber\\
&=\sqrt{2} \quad \textnormal{a.s.} \label{lawLog}
\end{align}
As a result we can rewrite~\eqref{klExpan} and conclude that
\begin{align*}
\MoveEqLeft[1]
\limsup	_{n\to\infty}\sqrt{n-k}\, D(\co{\hat P_{X_1^{k+1} Y_1^{k+1} Z_1^{k+1}}}{\bar P_{X_1^{k+1} Y_1^{k+1} Z_1^{k+1}}}) \\
&=\limsup_{n\to\infty} \frac{\log\log (n-k)}{\sqrt{n-k}} \sum\nolimits_{x_1^{k+1} y_1^{k+1} z_1^{k+1}}2C =0,
\end{align*}
given that each term in the finite sum is bounded.
Therefore, as $n\to \infty$, the four KL-divergence terms in~\eqref{DI_expansion} multiplied by a $\sqrt{n-k}$ factor tend to zero and the proof of Theorem~\ref{Convergence2_theo} is thus complete.
\endIEEEproof

\end{appendices}

% \addtolength{\textheight}{-21.7cm}

\bibliographystyle{IEEEtran}
\bibliography{IEEEabrv,ref}

\end{document}